\documentclass[preprint,apl,aip,superscriptaddress]{revtex4-2}
\usepackage[T2A]{fontenc}
\usepackage[utf8]{inputenc}
\usepackage[russian,english]{babel}
\usepackage{amsmath,amssymb,mathtools}
\usepackage{graphicx}
\usepackage{hyperref}

\begin{document}

\selectlanguage{english}

\title{Alignment and Timing Jitter in Flying-Focus Inverse Compton Scattering}
\author{S.G. Rykovanov}
\affiliation{Skolkovo Institute of Science and Technology, Moscow, Russia}
\affiliation{N.L. Dukhov All-Russian Research Institute of Automation, Moscow, Russia}
\email{serge.rykovanov@gmail.com}
\author{I. Dubinkin}
\affiliation{Skolkovo Institute of Science and Technology, Moscow, Russia}
\author{M.P. Malakhov}
\affiliation{Skolkovo Institute of Science and Technology, Moscow, Russia}
\affiliation{National Research Nuclear University MEPhI, Moscow, Russia}
\author{A.M. Fedotov}
\affiliation{National Research Nuclear University MEPhI, Moscow, Russia}
\author{A.S. Samsonov}
\affiliation{Institute of Applied Physics RAS, Nizhny Novgorod, Russia}
\author{I.Yu. Kostyukov}
\affiliation{Institute of Applied Physics RAS, Nizhny Novgorod, Russia}
\date{\today}

\begin{abstract}
Flying-focus laser pulses can extend the effective interaction length in inverse Compton sources by controlling the trajectory of the focal intensity. Their practical advantage, however, depends on tolerance to shot-to-shot electron--laser alignment and synchronization errors. We develop a semi-analytical model for the shot-averaged total photon yield in head-on inverse Compton scattering of an axisymmetric Gaussian electron bunch with a flying-focus laser pulse. The model includes finite electron-beam emittance, laser diffraction, transverse laser-centroid jitter, longitudinal focus-position jitter, and laser arrival-time jitter at the nominal interaction point. The ensemble averaging over these independent Gaussian errors and the integration over the longitudinal electron and laser coordinates are performed analytically, reducing the overlap problem to a single positive numerical quadrature. This formulation enables rapid evaluation of jitter-robust operating points and provides a compact tool for defining alignment and synchronization tolerances in flying-focus inverse Compton sources.

\end{abstract}
\maketitle

Inverse Compton scattering of laser photons by relativistic electron beams
provides compact, tunable x-ray and gamma-ray sources
\cite{KrafftPriebe2010,Albert2010,Graves2014,RykovanovJPhysB2014,INOK2023}.
Increasing the laser intensity raises the photon yield but also produces
nonlinear and ponderomotive spectral broadening, which can be important even
in the weakly nonlinear regime when a narrow bandwidth is required
\cite{Hartemann2010,RykovanovNCPM2025}.
Frequency modulation, temporal shaping, spectral caustics, and polarization
gating have been proposed to mitigate this broadening
\cite{SeiptPRA2015,RykovanovPRAB2016,TimoshenkoPRA2025,
KharinPRA2016,KharinPRL2018,SeiptPRL2019,ValialshchikovPRL2021}.
Here we consider a complementary geometric approach: increasing the
electron--laser overlap at fixed laser energy and controlled normalized
amplitude.

Flying-focus pulses provide a complementary geometric means of extending
laser--electron interactions. By controlling the trajectory of the focal
intensity independently of the laser envelope, a tightly focused region can
remain near a relativistic electron bunch over distances exceeding the
Rayleigh range \cite{Froula2018,Froula2019,Palastro2018}. Previous studies
have already established that this extended interaction can enhance Thomson
and inverse-Compton photon production, reduce spectral broadening, and
increase radiation-reaction and quantum-emission signatures
\cite{Ye2023,Formanek2022,Formanek2025}. Thus, the possibility of increasing
the ideal radiation yield with a flying focus is not itself the question
addressed here.

The practical gain depends on tolerance to transverse misalignment,
longitudinal focus-position jitter, and arrival-time jitter. These errors
affect the overlap through different mechanisms and cannot generally be
represented by a single longitudinal offset. We derive a semi-analytical
model including finite electron emittance, laser diffraction, and all three
Gaussian jitter sources. The ensemble-averaged yield reduces to a
one-dimensional integral, enabling rapid searches for jitter-robust focal
sizes and flying-focus velocities.

We consider a head-on collision between an electron bunch propagating along the positive $z$ direction and a flying-focus laser pulse propagating along the negative $z$ direction. The nominal electron waist and the ideal collision point are located at $x=y=z=t=0$. In the geometrical overlap model the electrons are assumed to be ultrarelativistic, so that their longitudinal velocity is approximately $\beta_e=v_e/c\simeq1$.

The electron bunch is assumed to be free of shot-to-shot fluctuations. The laser pulse is subject to three independent errors: a transverse centroid displacement $\Delta\mathbf{R}=(\Delta x,\Delta y)$, a longitudinal focus-position parameter $\Delta Z$, and an arrival-time error $\delta t$ at the nominal interaction point. The parameter $\Delta Z$ displaces the flying-focus trajectory without shifting the longitudinal laser envelope, whereas $\delta t$ shifts the laser envelope and the temporal origin of the moving focal structure.

For the axisymmetric beams considered here we introduce the transverse position vector $\mathbf{r}_{\perp}=(x,y)$. The electron bunch is described by
\begin{equation}
\begin{aligned}
n_e(\mathbf{r},t)
&=
\frac{N_e}
{(2\pi)^{3/2}\sigma_e^2(z)\sigma_{l,e}}
\exp\left[
-\frac{\left|\mathbf{r}_{\perp}\right|^2}
{2\sigma_e^2(z)}
-\frac{(z-ct)^2}
{2\sigma_{l,e}^2}
\right],
\end{aligned}
\label{eq:electron_density}
\end{equation}
where $N_e$ is the total number of electrons and $\sigma_{l,e}$ is the rms bunch length. The finite electron emittance is included through the beta-function evolution
\begin{equation}
\sigma_e^2(z)
=
\sigma_{e,0}^2
\left[
1+
\frac{z^2}
{\left(\mathcal{B}_e^*\right)^2}
\right],
\label{eq:electron_size}
\end{equation}
with
\begin{equation}
\sigma_{e,0}^2
=
\varepsilon_e\mathcal{B}_e^*,
\qquad
\varepsilon_e
=
\frac{\varepsilon_{n,e}}
{\gamma_e\beta_e}
\simeq
\frac{\varepsilon_{n,e}}
{\gamma_e}.
\label{eq:electron_emittance}
\end{equation}
Here $\mathcal{B}_e^*$ is the electron beta function at the waist, $\varepsilon_e$ is the geometric emittance, $\varepsilon_{n,e}$ is the normalized emittance, and $\gamma_e$ is the mean Lorentz factor.

The photon density of the flying-focus laser pulse is written as~\cite{Samsonov2026}
\begin{widetext}
\begin{equation}
\begin{aligned}
n_p^{\mathrm{FF}}
(\mathbf{r},t;\Delta\mathbf{R},\Delta Z,\delta t)
&=
\frac{N_p}
{(2\pi)^{3/2}
\sigma_p^2(z,t;\Delta Z,\delta t)
\sigma_{l,p}}
\exp\left[
-\frac{
\left|\mathbf{r}_{\perp}-\Delta\mathbf{R}\right|^2
}{
2\sigma_p^2(z,t;\Delta Z,\delta t)
}
-\frac{
\left[z+c\left(t-\delta t\right)\right]^2
}{
2\sigma_{l,p}^2
}
\right],
\end{aligned}
\label{eq:laser_density}
\end{equation}
\end{widetext}
where $N_p$ is the total number of laser photons and $\sigma_{l,p}$ is the rms pulse length. A positive $\delta t$ corresponds to a later passage of the laser-envelope center through the nominal interaction point $z=0$.

The rms laser spot size is modeled as
\begin{equation}
\sigma_p^2(z,t;\Delta Z,\delta t)
=
\sigma_{p,0}^2
\left[
1+
\frac{
\zeta^2(z,t;\Delta Z,\delta t)
}{
Z_R^2
}
\right],
\label{eq:laser_size}
\end{equation}
where $\sigma_{p,0}$ is the rms spot size at the moving waist and
\begin{equation}
Z_R
=
\frac{4\pi\sigma_{p,0}^2}{\lambda_0}
\label{eq:rayleigh_length}
\end{equation}
is the Rayleigh length for the central laser wavelength $\lambda_0$. The longitudinal coordinate relative to the moving waist is taken as~\cite{Samsonov2026}
\begin{equation}
\zeta(z,t;\Delta Z,\delta t)
=
\frac{
z-\beta_{\mathrm{ff}}c\left(t-\delta t\right)
}{
1+\beta_{\mathrm{ff}}
}
-\Delta Z,
\label{eq:zeta_definition}
\end{equation}
where $\beta_{\mathrm{ff}}=v_{\mathrm{ff}}/c$ is the flying-focus velocity. The instantaneous waist is located at $\zeta=0$ and follows
\begin{equation}
z_f(t)
=
\beta_{\mathrm{ff}}c\left(t-\delta t\right)
+
\left(1+\beta_{\mathrm{ff}}\right)\Delta Z.
\label{eq:waist_trajectory}
\end{equation}
The center of the longitudinal laser envelope follows
\begin{equation}
z_{\mathrm{env}}(t)
=
-c\left(t-\delta t\right).
\label{eq:envelope_trajectory}
\end{equation}
Consequently, $\delta t$ is the arrival-time error of the laser-envelope center at the nominal interaction point, rather than at the displaced focus. The envelope center and the moving waist intersect at
\begin{equation}
z_{\mathrm{int}}=\Delta Z,
\qquad
t_{\mathrm{int}}=\delta t-\frac{\Delta Z}{c}.
\label{eq:actual_focus_event}
\end{equation}
At fixed laboratory time, the waist-position displacement associated with $\Delta Z$ is $\left(1+\beta_{\mathrm{ff}}\right)\Delta Z$. For $\beta_{\mathrm{ff}}=0$, the model reduces to a stationary Gaussian waist displaced by $\Delta Z$, while the timing error shifts only the longitudinal laser envelope.

For a given realization of the alignment and arrival-time errors, the total number of scattered photons is obtained from the density-overlap integral
\begin{equation}
\begin{aligned}
N_\gamma(\Delta\mathbf{R},\Delta Z,\delta t)
&=
2c\sigma_{\mathrm{eff}}
\int_{-\infty}^{\infty}dt
\int d^3\mathbf{r}\,
n_e(\mathbf{r},t)
\\
&\quad\times
n_p^{\mathrm{FF}}
(\mathbf{r},t;\Delta\mathbf{R},\Delta Z,\delta t).
\end{aligned}
\label{eq:total_yield}
\end{equation}
Here $2c$ is the collinear M{\o}ller velocity~\cite{Moller1932}. In the Thomson regime $\sigma_{\mathrm{eff}}=\sigma_T$; when recoil is relevant, $\sigma_{\mathrm{eff}}$ is replaced by the total Klein--Nishina cross section evaluated at the mean collision energy.

We introduce the light-cone variables $\xi_e=z-ct$, $\xi_p=z+ct$, $\tau=c\delta t$. In these variables the coordinate relative to the moving waist becomes
\begin{equation}
\zeta(\xi_p,\xi_e;\Delta Z,\tau)
=
h_{\mathrm{ff}}\xi_p
+
\frac{\xi_e}{2}
+
g_{\mathrm{ff}}\tau
-
\Delta Z,
\label{eq:zeta_slice_coordinates}
\end{equation}
with
\begin{equation}
h_{\mathrm{ff}}
=
\frac{1-\beta_{\mathrm{ff}}}
{2(1+\beta_{\mathrm{ff}})},
\qquad
g_{\mathrm{ff}}
=
\frac{\beta_{\mathrm{ff}}}
{1+\beta_{\mathrm{ff}}}.
\label{eq:ff_coefficients}
\end{equation}
The timing error therefore shifts the laser envelope through $\xi_p-\tau$ and, for $\beta_{\mathrm{ff}}\neq0$, also shifts the moving waist through the term $g_{\mathrm{ff}}\tau$.

For compactness we define
\begin{equation}
\kappa_e
=
\frac{\sigma_{e,0}^2}{(\mathcal{B}_e^*)^2},
\qquad
\kappa_p
=
\frac{\sigma_{p,0}^2}{Z_R^2},
\qquad
S_0
=
\sigma_{e,0}^2+\sigma_{p,0}^2.
\label{eq:kappa_definitions}
\end{equation}
The combined transverse variance is then
\begin{equation}
\begin{aligned}
S(\xi_p,\xi_e;\Delta Z,\tau)
&=
S_0
+
\kappa_e
\left(
\frac{\xi_p+\xi_e}{2}
\right)^2
\\
&\quad+
\kappa_p
\left(
h_{\mathrm{ff}}\xi_p
+
\frac{\xi_e}{2}
+
g_{\mathrm{ff}}\tau
-
\Delta Z
\right)^2.
\end{aligned}
\label{eq:combined_width}
\end{equation}

After the transverse Gaussian integration, Eq.~\eqref{eq:total_yield} becomes
\begin{widetext}
\begin{equation}
\begin{aligned}
N_\gamma(\Delta\mathbf{R},\Delta Z,\tau)
&=
\frac{
\sigma_{\mathrm{eff}}N_eN_p
}{
(2\pi)^2\sigma_{l,e}\sigma_{l,p}
}
\int_{-\infty}^{\infty}d\xi_p
\int_{-\infty}^{\infty}d\xi_e
\frac{
\exp\left[
-\frac{(\xi_p-\tau)^2}{2\sigma_{l,p}^2}
-\frac{\xi_e^2}{2\sigma_{l,e}^2}
-\frac{|\Delta\mathbf{R}|^2}
{2S(\xi_p,\xi_e;\Delta Z,\tau)}
\right]
}{
S(\xi_p,\xi_e;\Delta Z,\tau)
}.
\end{aligned}
\label{eq:total_yield_full}
\end{equation}
\end{widetext}

We assume independent, zero-mean Gaussian transverse centroid, longitudinal focus-position, and arrival-time jitter. Their rms values are defined as $\left\langle \Delta x^2 \right\rangle^{1/2}=\left\langle \Delta y^2 \right\rangle^{1/2}\equiv\sigma_{J,\perp}$, $\left\langle \Delta Z^2 \right\rangle^{1/2}\equiv\sigma_{J,Z}$, $\left\langle \tau^2 \right\rangle^{1/2}=c\sigma_{J,t}\equiv\sigma_{J,\tau}$.
The transverse average of Eq.~\eqref{eq:total_yield_full} is analytic and amounts to the replacement $S(\xi_p,\xi_e;\Delta Z,\tau)\rightarrow\widetilde S(\xi_p,\xi_e;\Delta Z,\tau)=S(\xi_p,\xi_e;\Delta Z,\tau)+\sigma_{J,\perp}^2$.

The remaining averages over $\Delta Z$ and $\tau$ are performed after introducing the Laplace representation
\begin{equation}
\frac{1}{\widetilde S}
=
\int_0^\infty ds\,
\exp\left(-s\widetilde S\right),
\label{eq:laplace_representation}
\end{equation}
with $\widetilde S_0=S_0+\sigma_{J,\perp}^2$. The integrations over $\Delta Z$, $\xi_p$, $\xi_e$, and $\tau$ are then Gaussian. The shot-averaged total yield reduces to
\begin{equation}
\begin{aligned}
\left\langle N_\gamma \right\rangle
&=
\frac{
\sigma_{\mathrm{eff}}N_eN_p
}{
2\pi\sigma_{l,e}\sigma_{l,p}
}
\int_0^\infty ds\,
\frac{
\exp\left(-s\widetilde S_0\right)
}{
\sqrt{\mathcal Q(s)}
},
\end{aligned}
\label{eq:master_integral}
\end{equation}
where the complete dependence on the Laplace variable is contained in the quadratic polynomial
\begin{equation}
\mathcal Q(s)
=
q_0+q_1s+q_2s^2.
\label{eq:Q_polynomial}
\end{equation}
To write the coefficients compactly, we introduce
\begin{equation}
\alpha_p
=
\frac{1}{\sigma_{l,p}^2},
\qquad
\alpha_e
=
\frac{1}{\sigma_{l,e}^2},
\qquad
\chi_Z
=
2\kappa_p\sigma_{J,Z}^2.
\label{eq:master_auxiliary_coefficients}
\end{equation}
The coefficients of $\mathcal Q(s)$ are then
\begin{equation}
q_0
=
\alpha_p\alpha_e,
\label{eq:q0}
\end{equation}
\begin{equation}
\begin{aligned}
q_1
&=
\chi_Z\alpha_p\alpha_e
+
\frac{\kappa_e}{2}
\left(
\sigma_{J,\tau}^2\alpha_p\alpha_e
+
\alpha_p+\alpha_e
\right)
\\
&\quad+
\frac{\kappa_p}{2}
\left(
\sigma_{J,\tau}^2\alpha_p\alpha_e
+
\alpha_p
+
4h_{\mathrm{ff}}^2\alpha_e
\right),
\end{aligned}
\label{eq:q1}
\end{equation}
and
\begin{equation}
\begin{aligned}
q_2
&=
\frac{\chi_Z\kappa_e}{2}
\left(
\sigma_{J,\tau}^2\alpha_p\alpha_e
+
\alpha_p+\alpha_e
\right)
\\
&\quad+
\kappa_e\kappa_p g_{\mathrm{ff}}^2
\left(
1+\sigma_{J,\tau}^2\alpha_e
\right).
\end{aligned}
\label{eq:q2}
\end{equation}
Transverse centroid jitter enters through $\widetilde S_0$, whereas
longitudinal focus-position and arrival-time jitter enter through $\chi_Z$
and $\sigma_{J,\tau}$, respectively. The polynomial representation remains
regular for $\sigma_{J,t}\rightarrow0$. Its derivation from the underlying
Gaussian integrals and determinant representation is given in the
Supplemental Material.

\begin{figure}[t]
\centering
\includegraphics[width=\columnwidth]{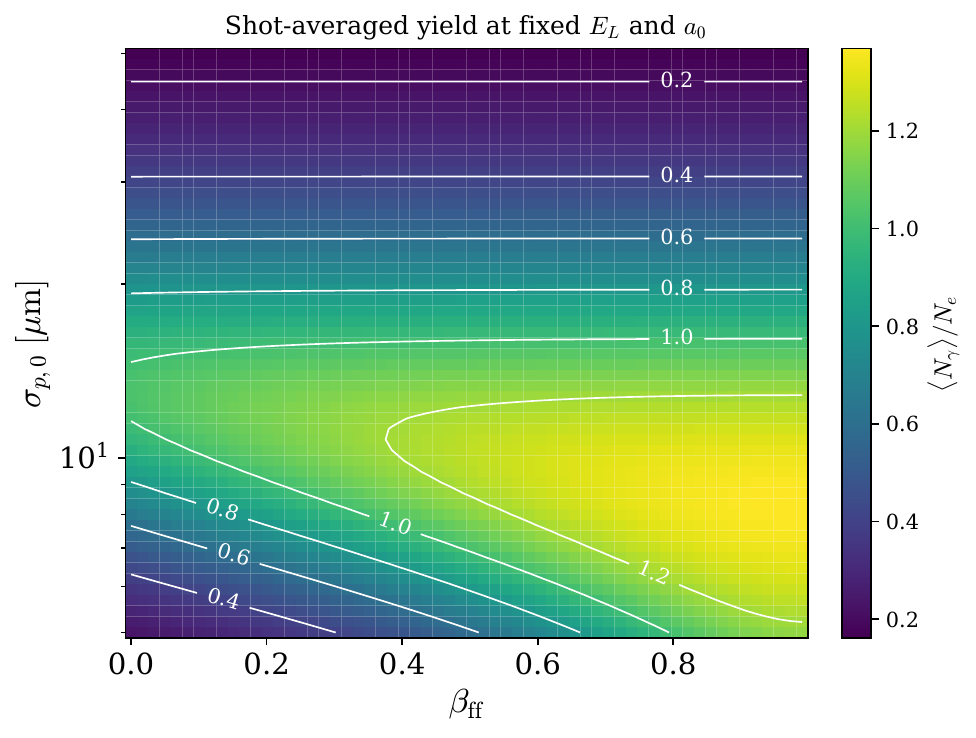}
\caption{
\label{fig:spot_beta_map}
Shot-averaged photon yield per electron as a function of the rms laser waist
$\sigma_{p,0}$ and flying-focus velocity $\beta_{\mathrm{ff}}$ for the beam
parameters specified in the text. Independent Gaussian jitter is included
with $\sigma_{J,\perp}=6.54~\mu\mathrm{m}$,
$\sigma_{J,Z}=2.5~\mathrm{mm}$, and
$\sigma_{J,t}=10~\mathrm{ps}$. White curves are equal-yield contours.
The Klein--Nishina cross section is used.
}
\end{figure}

We first use Eq.~\eqref{eq:master_integral} to map the jitter-averaged photon yield as a function of the nominal laser waist and the flying-focus velocity. Unless stated otherwise, the electron bunch has kinetic energy $E_e=1~\mathrm{GeV}$, normalized emittance $\varepsilon_{n,e}=1~\mathrm{mm\,mrad}$, beta function $\mathcal{B}_e^*=50~\mathrm{mm}$, and rms length $\sigma_{l,e}=c\times1~\mathrm{ps}\approx300~\mu\mathrm{m}$. The laser wavelength, energy, and peak normalized amplitude are fixed at $\lambda_0=0.8~\mu\mathrm{m}$, $E_L=10~\mathrm{J}$, and $a_0=0.1$. Keeping $a_0$ fixed isolates the geometric effect of the flying focus and prevents variations in nonlinear broadening across the scan. A related analysis for an ordinary Gaussian laser pulse without a flying focus was presented in Ref.~\cite{RykovanovJPhysB2014}.

Figure~\ref{fig:spot_beta_map} shows the shot-averaged photon yield when transverse centroid jitter, longitudinal focus-position jitter, and arrival-time jitter at the nominal interaction point are included simultaneously. The scan is performed at fixed laser energy, wavelength, and peak normalized amplitude.  At fixed $E_L$ and $a_0$, the rms laser pulse length scales as $\sigma_{l,p}\propto\sigma_{p,0}^{-2}$. Therefore, tighter focusing is accompanied by a longer pulse. For a stationary waist, $\beta_{\mathrm{ff}}=0$, a substantial part of such a pulse interacts away from the focal region because of diffraction. A nonzero flying-focus velocity allows the waist to remain close to the electron bunch over a larger fraction of the collision, increasing the effective interaction length. The map in Fig.~\ref{fig:spot_beta_map} therefore represents a simultaneous tolerance test against transverse misalignment, longitudinal focus-position jitter, and synchronization error. The persistence of an enhanced-yield region demonstrates that flying-focus control can improve the shot-averaged, rather than only ideal, inverse-Compton yield under realistic alignment and timing fluctuations.

For the beam, laser, and jitter parameters specified above, a continuous optimization of Eq.~\eqref{eq:master_integral} gives a jitter-robust operating point at
$(\sigma_{p,0}^{\mathrm{opt}},\beta_{\mathrm{ff}}^{\mathrm{opt}})
=(8.36~\mu\mathrm{m},1.00)$, where
$\langle N_\gamma\rangle/N_e=1.37$.
At the same laser waist and jitter amplitudes, a stationary focus gives
$\langle N_\gamma\rangle/N_e=0.703$, corresponding to a flying-focus
enhancement factor of $1.95$. If the stationary-focus waist is optimized
independently for the same beam and laser parameters, its maximum yield is
$1.03$ photons per electron at $\sigma_{p,0}=13.0~\mu\mathrm{m}$, giving a
more conservative design-to-design enhancement factor of $1.33$.

For the same beam and laser parameters but in the absence of jitter, the
optimum shifts to
$(\sigma_{p,0}^{\mathrm{opt}},\beta_{\mathrm{ff}}^{\mathrm{opt}})
=(3.90~\mu\mathrm{m},1.00)$, with
$\langle N_\gamma\rangle/N_e=7.92$.
The corresponding optimized stationary-focus yield is $1.45$ photons per
electron at $\sigma_{p,0}=13.0~\mu\mathrm{m}$, yielding an ideal
flying-focus enhancement factor of $5.45$. The combined alignment and timing
jitter therefore broadens the optimal waist by a factor of approximately
$2.1$ and reduces, but does not eliminate, the flying-focus advantage.

\begin{figure*}[t]
\centering
\includegraphics[width=\textwidth]{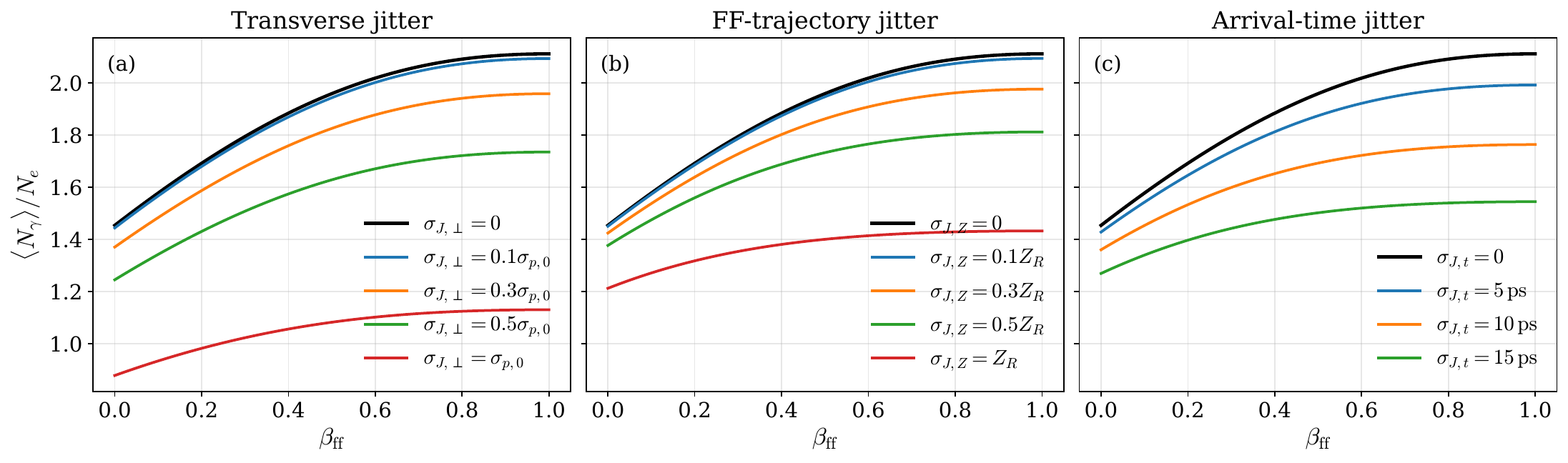}
\caption{
\label{fig:three_jitter_panels}
Shot-averaged photon yield per electron versus $\beta_{\mathrm{ff}}$ for
independently varied (a) transverse centroid jitter, (b) longitudinal
focus-position jitter, and (c) arrival-time jitter. The other two jitter
sources are set to zero in each panel, and the black curve is the ideal
no-jitter result. Beam parameters are the same as in
Fig.~\ref{fig:spot_beta_map}.
}
\end{figure*}

To isolate the effects of the three error sources, Fig.~\ref{fig:three_jitter_panels} shows the shot-averaged yield when transverse centroid jitter, longitudinal focus-position jitter, and arrival-time jitter are varied independently.

The three jitter sources produce distinct responses. Transverse jitter
mainly reduces the overall overlap, while focus-position jitter modifies the
diffraction-limited matching between the moving waist and the electron
bunch. Arrival-time jitter shifts both the laser envelope and, for
$\beta_{\mathrm{ff}}\neq0$, the waist position at a given laboratory time;
it therefore cannot be absorbed into an effective $\sigma_{J,Z}$. In all
three cases the favorable finite-$\beta_{\mathrm{ff}}$ region survives over
a substantial range of jitter amplitudes, although its gain decreases as
the fluctuations increase.

In conclusion, we derived a semi-analytical expression for the
shot-averaged photon yield in flying-focus inverse Compton scattering,
including finite electron emittance, laser diffraction, transverse
centroid jitter, longitudinal focus-position jitter, and arrival-time
jitter. All spatial and temporal integrations reduce to a single positive
quadrature. The calculations show that a finite flying-focus velocity can
increase the jitter-averaged yield and that the three error sources affect
the overlap through distinct mechanisms. The model therefore provides a
rapid tool for identifying robust operating points and setting alignment
and synchronization tolerances.

This work was funded within the National Center for Physics and Mathematics (NCPhM, Sarov) direction 6. The data and code required to reproduce all figures are openly available in
the \texttt{compton\_ff\_yield} repository
\cite{RykovanovComptonFFYield2026}.

\section{Appendix: Derivation of the shot-averaged overlap integral}
\label{app:derivation}

This Appendix gives the derivation of the one-dimensional master integral used in the main text. We consider a head-on collision between an electron bunch propagating along the positive $z$ direction and a flying-focus laser pulse propagating along the negative $z$ direction. The nominal electron waist and the ideal interaction point are located at $x=y=z=t=0$. The electron beam is assumed to be free of shot-to-shot fluctuations. The laser pulse is subject to three mutually independent, zero-mean Gaussian errors: a transverse centroid displacement $\Delta\mathbf R=(\Delta x,\Delta y)$, a longitudinal focus-position displacement $\Delta Z$, and an arrival-time error $\delta t$ defined at the nominal interaction point.

The electron density is
\begin{equation}
\begin{aligned}
n_e(\mathbf r,t)
&=
\frac{N_e}
{(2\pi)^{3/2}\sigma_e^2(z)\sigma_{l,e}}
\exp\left[
-\frac{|\mathbf r_\perp|^2}
{2\sigma_e^2(z)}
-\frac{(z-ct)^2}
{2\sigma_{l,e}^2}
\right],
\end{aligned}
\label{app:eq:electron_density}
\end{equation}
where $\mathbf r_\perp=(x,y)$, $N_e$ is the number of electrons, and $\sigma_{l,e}$ is the rms electron bunch length. The finite emittance of the electron beam is included through
\begin{equation}
\sigma_e^2(z)
=
\sigma_{e,0}^2
\left[
1+
\frac{z^2}
{(\mathcal B_e^*)^2}
\right],
\label{app:eq:electron_size}
\end{equation}
with
\begin{equation}
\sigma_{e,0}^2
=
\varepsilon_e\mathcal B_e^*,
\qquad
\varepsilon_e
=
\frac{\varepsilon_{n,e}}
{\gamma_e\beta_e}
\simeq
\frac{\varepsilon_{n,e}}
{\gamma_e}.
\label{app:eq:electron_emittance}
\end{equation}

The photon density of the flying-focus laser pulse is written as
\begin{equation}
\begin{aligned}
n_p^{\mathrm{FF}}
(\mathbf r,t;\Delta\mathbf R,\Delta Z,\delta t)
&=
\frac{N_p}
{(2\pi)^{3/2}
\sigma_p^2(z,t;\Delta Z,\delta t)
\sigma_{l,p}}
\\
&\quad\times
\exp\left[
-\frac{
|\mathbf r_\perp-\Delta\mathbf R|^2
}
{
2\sigma_p^2(z,t;\Delta Z,\delta t)
}
-\frac{
[z+c(t-\delta t)]^2
}
{
2\sigma_{l,p}^2
}
\right],
\end{aligned}
\label{app:eq:laser_density}
\end{equation}
where $N_p$ is the number of laser photons and $\sigma_{l,p}$ is the rms laser pulse length. At every fixed time, Eqs.~\eqref{app:eq:electron_density} and \eqref{app:eq:laser_density} are normalized to $N_e$ and $N_p$, respectively. A positive $\delta t$ corresponds to a later laser arrival at the nominal interaction point. The transverse rms size of the laser pulse is modeled as
\begin{equation}
\sigma_p^2(z,t;\Delta Z,\delta t)
=
\sigma_{p,0}^2
\left[
1+
\frac{
\zeta^2(z,t;\Delta Z,\delta t)
}
{Z_R^2}
\right],
\label{app:eq:laser_size}
\end{equation}
where
\begin{equation}
Z_R
=
\frac{4\pi\sigma_{p,0}^2}{\lambda_0}
\label{app:eq:rayleigh_length}
\end{equation}
is the Rayleigh length and
\begin{equation}
\zeta(z,t;\Delta Z,\delta t)
=
\frac{
z-\beta_{\mathrm{ff}}c(t-\delta t)
}
{1+\beta_{\mathrm{ff}}}
-\Delta Z.
\label{app:eq:zeta_definition}
\end{equation}
Here $\beta_{\mathrm{ff}}=v_{\mathrm{ff}}/c$ is the flying-focus velocity. The instantaneous waist satisfies $\zeta=0$ and therefore follows
\begin{equation}
z_f(t)
=
\beta_{\mathrm{ff}}c(t-\delta t)
+
(1+\beta_{\mathrm{ff}})\Delta Z.
\label{app:eq:waist_trajectory}
\end{equation}
Thus, at fixed time, the physical waist-position shift associated with $\Delta Z$ is $(1+\beta_{\mathrm{ff}})\Delta Z$.

The center of the longitudinal laser envelope follows
\begin{equation}
z_{\mathrm{env}}(t)
=
-c(t-\delta t).
\label{app:eq:envelope_trajectory}
\end{equation}
Consequently, $\delta t$ denotes the arrival-time error of the laser-envelope center at the nominal interaction point $z=0$, rather than its arrival time at the displaced focus. The envelope center and the moving waist intersect at the actual focus event
\begin{equation}
z_*=\Delta Z,
\qquad
t_*=\delta t-\frac{\Delta Z}{c}.
\label{app:eq:actual_focus_event}
\end{equation}
Thus, $\Delta Z$ determines the longitudinal position of the actual focus event, whereas $\delta t$ determines the arrival time of the laser-envelope center at the nominal interaction point.

For two relativistic particle populations, the collision-rate flux factor is the M{\o}ller velocity \cite{Moller1932},
\begin{equation}
v_{\mathrm M}
=
\sqrt{
\left|\mathbf v_e-\mathbf v_p\right|^2
-
\frac{
\left|\mathbf v_e\times\mathbf v_p\right|^2
}{c^2}
}.
\label{app:eq:moller_velocity}
\end{equation}
For the head-on geometry,
$\mathbf v_e=\beta_ec\,\hat{\mathbf z}$ and
$\mathbf v_p=-c\,\hat{\mathbf z}$, so that
\begin{equation}
v_{\mathrm M}
=
c(1+\beta_e)
\simeq
2c.
\label{app:eq:moller_headon}
\end{equation}
The M{\o}ller velocity is a relativistic flux factor and should not be interpreted as the physical speed of one particle measured in the rest frame of the other.

Using the ultrarelativistic approximation $v_{\mathrm M}\simeq2c$, the total number of scattered photons for a given realization of the errors is
\begin{equation}
\begin{aligned}
N_\gamma(\Delta\mathbf R,\Delta Z,\delta t)
&=
2c\sigma_{\mathrm{eff}}
\int_{-\infty}^{\infty}dt
\int d^3\mathbf r\,
n_e(\mathbf r,t)
\\
&\quad\times
n_p^{\mathrm{FF}}
(\mathbf r,t;\Delta\mathbf R,\Delta Z,\delta t).
\end{aligned}
\label{app:eq:total_yield}
\end{equation}
The derivation assumes that $\sigma_{\mathrm{eff}}$ can be factored out of the space--time integral. In the Thomson regime $\sigma_{\mathrm{eff}}=\sigma_T$; for sufficiently narrow electron-energy and laser-frequency distributions, recoil can be included by evaluating the total Klein--Nishina cross section at the mean collision energy. Electron-energy evolution during repeated scattering events is not included.

\subsection{Transverse Gaussian integration}

For fixed $z$ and $t$, define the combined transverse variance
\begin{equation}
S(z,t;\Delta Z,\delta t)
=
\sigma_e^2(z)
+
\sigma_p^2(z,t;\Delta Z,\delta t).
\label{app:eq:S_definition}
\end{equation}
The transverse part of the overlap is
\begin{equation}
\begin{aligned}
I_\perp
&=
\int d^2\mathbf r_\perp\,
\frac{1}{2\pi\sigma_e^2}
\exp\left[
-\frac{|\mathbf r_\perp|^2}{2\sigma_e^2}
\right]
\\
&\quad\times
\frac{1}{2\pi\sigma_p^2}
\exp\left[
-\frac{|\mathbf r_\perp-\Delta\mathbf R|^2}{2\sigma_p^2}
\right].
\end{aligned}
\label{app:eq:transverse_overlap_start}
\end{equation}
Completing the square gives the standard two-dimensional Gaussian convolution
\begin{equation}
I_\perp
=
\frac{1}{2\pi S}
\exp\left[
-\frac{|\Delta\mathbf R|^2}{2S}
\right].
\label{app:eq:transverse_overlap_result}
\end{equation}
This is the origin of the factor $S^{-1}\exp[-|\Delta\mathbf R|^2/(2S)]$ in the reduced overlap integral.

\subsection{Light-cone variables}

We introduce
\begin{equation}
\xi_e=z-ct,
\qquad
\xi_p=z+ct,
\qquad
\tau=c\delta t.
\label{app:eq:light_cone_variables}
\end{equation}
The inverse transformation is
\begin{equation}
z=
\frac{\xi_p+\xi_e}{2},
\qquad
ct=
\frac{\xi_p-\xi_e}{2},
\qquad
2c\,dz\,dt=d\xi_p\,d\xi_e.
\label{app:eq:inverse_light_cone}
\end{equation}
The longitudinal electron and laser envelopes become
\begin{equation}
z-ct=\xi_e,
\qquad
z+c(t-\delta t)=\xi_p-\tau.
\label{app:eq:longitudinal_envelopes}
\end{equation}
The electron transverse size becomes
\begin{equation}
\sigma_e^2(\xi_p,\xi_e)
=
\sigma_{e,0}^2
\left[
1+
\frac{(\xi_p+\xi_e)^2}
{4(\mathcal B_e^*)^2}
\right].
\label{app:eq:electron_size_light_cone}
\end{equation}
The coordinate relative to the moving waist becomes
\begin{equation}
\zeta(\xi_p,\xi_e;\Delta Z,\tau)
=
h_{\mathrm{ff}}\xi_p
+
\frac{\xi_e}{2}
+
g_{\mathrm{ff}}\tau
-
\Delta Z,
\label{app:eq:zeta_light_cone}
\end{equation}
where
\begin{equation}
h_{\mathrm{ff}}
=
\frac{1-\beta_{\mathrm{ff}}}
{2(1+\beta_{\mathrm{ff}})},
\qquad
g_{\mathrm{ff}}
=
\frac{\beta_{\mathrm{ff}}}
{1+\beta_{\mathrm{ff}}}.
\label{app:eq:hg_definitions}
\end{equation}
For $\beta_{\mathrm{ff}}=0$, one has $h_{\mathrm{ff}}=1/2$ and $g_{\mathrm{ff}}=0$, so that $\zeta=z-\Delta Z$ and the timing error shifts only the laser envelope.

After the transverse integration and the light-cone transformation, Eq.~\eqref{app:eq:total_yield} becomes
\begin{equation}
\begin{aligned}
N_\gamma(\Delta\mathbf R,\Delta Z,\tau)
&=
\frac{
\sigma_{\mathrm{eff}}N_eN_p
}
{
(2\pi)^2\sigma_{l,e}\sigma_{l,p}
}
\int_{-\infty}^{\infty}d\xi_p
\int_{-\infty}^{\infty}d\xi_e
\\
&\quad\times
\frac{
\exp\left[
-\frac{(\xi_p-\tau)^2}{2\sigma_{l,p}^2}
-\frac{\xi_e^2}{2\sigma_{l,e}^2}
-\frac{|\Delta\mathbf R|^2}
{2S(\xi_p,\xi_e;\Delta Z,\tau)}
\right]
}
{
S(\xi_p,\xi_e;\Delta Z,\tau)
}.
\end{aligned}
\label{app:eq:fixed_shot_yield}
\end{equation}

For compactness, define
\begin{equation}
\kappa_e
=
\frac{\sigma_{e,0}^2}{(\mathcal B_e^*)^2},
\qquad
\kappa_p
=
\frac{\sigma_{p,0}^2}{Z_R^2},
\qquad
S_0
=
\sigma_{e,0}^2+\sigma_{p,0}^2.
\label{app:eq:kappa_definitions}
\end{equation}
Then
\begin{equation}
\begin{aligned}
S(\xi_p,\xi_e;\Delta Z,\tau)
&=
S_0
+
\kappa_e
\left(
\frac{\xi_p+\xi_e}{2}
\right)^2
\\
&\quad+
\kappa_p
\left(
h_{\mathrm{ff}}\xi_p
+
\frac{\xi_e}{2}
+
g_{\mathrm{ff}}\tau
-
\Delta Z
\right)^2.
\end{aligned}
\label{app:eq:S_light_cone}
\end{equation}

\subsection{Averaging over transverse centroid jitter}

The transverse centroid displacement is assumed to be Gaussian and isotropic,
\begin{equation}
P_\perp(\Delta\mathbf R)
=
\frac{1}{2\pi\sigma_{J,\perp}^2}
\exp\left[
-\frac{|\Delta\mathbf R|^2}
{2\sigma_{J,\perp}^2}
\right],
\label{app:eq:transverse_jitter_distribution}
\end{equation}
with
\begin{equation}
\left\langle \Delta x^2\right\rangle^{1/2}
=
\left\langle \Delta y^2\right\rangle^{1/2}
=
\sigma_{J,\perp}.
\label{app:eq:transverse_jitter_rms}
\end{equation}
The transverse jitter average of the factor containing $\Delta\mathbf R$ is
\begin{equation}
\begin{aligned}
&
\int d^2\Delta\mathbf R\,
P_\perp(\Delta\mathbf R)
\frac{1}{S}
\exp\left[
-\frac{|\Delta\mathbf R|^2}{2S}
\right]
\\
&=
\frac{1}{S}
\frac{1}{\sigma_{J,\perp}^2}
\int_0^\infty dR\,R
\exp\left[
-\frac{R^2}{2}
\left(
\frac{1}{S}
+
\frac{1}{\sigma_{J,\perp}^2}
\right)
\right]
\\
&=
\frac{1}{S+\sigma_{J,\perp}^2}.
\end{aligned}
\label{app:eq:transverse_jitter_average}
\end{equation}
Therefore transverse Gaussian centroid jitter enters as an additive contribution to the combined transverse variance:
\begin{equation}
\widetilde S(\xi_p,\xi_e;\Delta Z,\tau)
=
S(\xi_p,\xi_e;\Delta Z,\tau)
+
\sigma_{J,\perp}^2.
\label{app:eq:S_tilde}
\end{equation}
Equivalently,
\begin{equation}
\widetilde S_0
=
S_0+\sigma_{J,\perp}^2.
\label{app:eq:S0_tilde}
\end{equation}
The transverse-jitter-averaged yield at fixed $\Delta Z$ and $\tau$ is
\begin{equation}
\begin{aligned}
\left\langle
N_\gamma
\right\rangle_\perp
(\Delta Z,\tau)
&=
\frac{
\sigma_{\mathrm{eff}}N_eN_p
}
{
(2\pi)^2\sigma_{l,e}\sigma_{l,p}
}
\int_{-\infty}^{\infty}d\xi_p
\int_{-\infty}^{\infty}d\xi_e
\\
&\quad\times
\frac{
\exp\left[
-\frac{(\xi_p-\tau)^2}{2\sigma_{l,p}^2}
-\frac{\xi_e^2}{2\sigma_{l,e}^2}
\right]
}
{
\widetilde S(\xi_p,\xi_e;\Delta Z,\tau)
}.
\end{aligned}
\label{app:eq:transverse_averaged_yield}
\end{equation}

\subsection{Laplace representation}

The remaining denominator is treated with the Laplace representation
\begin{equation}
\frac{1}{\widetilde S}
=
\int_0^\infty ds\,
\exp(-s\widetilde S),
\qquad
\widetilde S>0.
\label{app:eq:laplace_representation}
\end{equation}
We define
\begin{equation}
K_e=s\kappa_e,
\qquad
K_p=s\kappa_p.
\label{app:eq:K_definitions}
\end{equation}
The longitudinal coordinate entering the laser diffraction term can be written as
\begin{equation}
u(\xi_p,\xi_e,\tau)
=
h_{\mathrm{ff}}\xi_p
+
\frac{\xi_e}{2}
+
g_{\mathrm{ff}}\tau.
\label{app:eq:u_definition}
\end{equation}
Then
\begin{equation}
\widetilde S
=
\widetilde S_0
+
\kappa_e
\left(
\frac{\xi_p+\xi_e}{2}
\right)^2
+
\kappa_p(u-\Delta Z)^2.
\label{app:eq:S_tilde_compact}
\end{equation}
After applying Eq.~\eqref{app:eq:laplace_representation}, the $\Delta Z$-dependent factor is
\begin{equation}
\exp\left[
-K_p(u-\Delta Z)^2
\right].
\label{app:eq:DeltaZ_factor}
\end{equation}

\subsection{Averaging over longitudinal focus-position jitter}

The longitudinal focus-position displacement is assumed to be Gaussian,
\begin{equation}
P_Z(\Delta Z)
=
\frac{1}{\sqrt{2\pi}\sigma_{J,Z}}
\exp\left[
-\frac{\Delta Z^2}
{2\sigma_{J,Z}^2}
\right],
\qquad
\left\langle \Delta Z^2 \right\rangle^{1/2}
=
\sigma_{J,Z}.
\label{app:eq:DeltaZ_distribution}
\end{equation}
According to Eq.~\eqref{app:eq:actual_focus_event}, $\sigma_{J,Z}$ is the rms longitudinal displacement of the actual envelope--waist intersection. At fixed laboratory time, the corresponding rms displacement of the waist trajectory is $(1+\beta_{\mathrm{ff}})\sigma_{J,Z}$. The required average is
\begin{equation}
\begin{aligned}
I_Z
&=
\int_{-\infty}^{\infty}
\frac{d\Delta Z}
{\sqrt{2\pi}\sigma_{J,Z}}
\exp\left[
-\frac{\Delta Z^2}{2\sigma_{J,Z}^2}
-K_p(u-\Delta Z)^2
\right].
\end{aligned}
\label{app:eq:IZ_start}
\end{equation}
Combining the quadratic terms gives
\begin{equation}
-\frac{\Delta Z^2}{2\sigma_{J,Z}^2}
-K_p(u-\Delta Z)^2
=
-\left(
K_p+\frac{1}{2\sigma_{J,Z}^2}
\right)\Delta Z^2
+
2K_pu\Delta Z
-
K_pu^2.
\label{app:eq:IZ_quadratic}
\end{equation}
The Gaussian integral can be evaluated by completing the square. The result is
\begin{equation}
I_Z
=
\frac{
\exp(-\Lambda_pu^2)
}
{
\sqrt{1+2K_p\sigma_{J,Z}^2}
},
\label{app:eq:IZ_result}
\end{equation}
where
\begin{equation}
\Lambda_p
=
\frac{K_p}
{1+2K_p\sigma_{J,Z}^2}.
\label{app:eq:Lambda_definition}
\end{equation}
Thus, the longitudinal focus-position jitter reduces the effective coefficient of the laser-diffraction quadratic form from $K_p$ to $\Lambda_p$ and produces the normalization factor
\begin{equation}
\left(
1+2K_p\sigma_{J,Z}^2
\right)^{-1/2}.
\label{app:eq:DeltaZ_normalization}
\end{equation}

\subsection{Including arrival-time jitter at the nominal interaction point}

The arrival-time-jitter variable is
\begin{equation}
\tau=c\delta t,
\qquad
\left\langle \tau^2 \right\rangle^{1/2}
=
\sigma_{J,\tau}
=
c\sigma_{J,t}.
\label{app:eq:tau_definition}
\end{equation}
Its Gaussian distribution is
\begin{equation}
P_\tau(\tau)
=
\frac{1}{\sqrt{2\pi}\sigma_{J,\tau}}
\exp\left[
-\frac{\tau^2}{2\sigma_{J,\tau}^2}
\right].
\label{app:eq:tau_distribution}
\end{equation}
After averaging over $\Delta Z$, the exponent depending on $\xi_p$, $\xi_e$, and $\tau$ is
\begin{equation}
\begin{aligned}
\Phi(\xi_p,\xi_e,\tau)
&=
\frac{(\xi_p-\tau)^2}{2\sigma_{l,p}^2}
+
\frac{\xi_e^2}{2\sigma_{l,e}^2}
+
\frac{K_e}{4}(\xi_p+\xi_e)^2
\\
&\quad+
\Lambda_p
\left(
h_{\mathrm{ff}}\xi_p
+
\frac{\xi_e}{2}
+
g_{\mathrm{ff}}\tau
\right)^2
+
\frac{\tau^2}{2\sigma_{J,\tau}^2}.
\end{aligned}
\label{app:eq:Phi_definition}
\end{equation}
It is useful to introduce the vector
\begin{equation}
\mathbf q
=
\begin{pmatrix}
\xi_p \\
\xi_e \\
\tau
\end{pmatrix}.
\label{app:eq:q_vector}
\end{equation}
Then
\begin{equation}
\Phi
=
\frac{1}{2}
\mathbf q^{\mathsf T}
\mathbf M(s)
\mathbf q.
\label{app:eq:Phi_matrix}
\end{equation}
The matrix $\mathbf M(s)$ is
\begin{equation}
\begin{aligned}
\mathbf M(s)
&=
\frac{1}{\sigma_{l,p}^2}
\mathbf v_p\mathbf v_p^{\mathsf T}
+
\frac{1}{\sigma_{l,e}^2}
\mathbf v_e\mathbf v_e^{\mathsf T}
+
\frac{K_e}{2}
\mathbf v_d\mathbf v_d^{\mathsf T}
\\
&\quad+
2\Lambda_p
\mathbf v_{\mathrm{ff}}\mathbf v_{\mathrm{ff}}^{\mathsf T}
+
\frac{1}{\sigma_{J,\tau}^2}
\mathbf v_\tau\mathbf v_\tau^{\mathsf T},
\end{aligned}
\label{app:eq:M_matrix}
\end{equation}
with
\begin{equation}
\begin{gathered}
\mathbf v_p=(1,0,-1)^{\mathsf T},
\qquad
\mathbf v_e=(0,1,0)^{\mathsf T},
\qquad
\mathbf v_d=(1,1,0)^{\mathsf T},
\\
\mathbf v_{\mathrm{ff}}
=
\left(
h_{\mathrm{ff}},
\frac{1}{2},
g_{\mathrm{ff}}
\right)^{\mathsf T},
\qquad
\mathbf v_\tau=(0,0,1)^{\mathsf T}.
\end{gathered}
\label{app:eq:M_vectors}
\end{equation}
The vector $\mathbf v_p$ represents the laser envelope through the combination $\xi_p-\tau$, whereas $\mathbf v_{\mathrm{ff}}$ represents the flying-focus diffraction coordinate. Therefore, timing jitter enters the longitudinal envelope and, when $g_{\mathrm{ff}}\neq0$, also the moving-waist coordinate.

Equivalently, the same matrix can be written component-wise as
\begin{equation}
\mathbf M
=
\begin{pmatrix}
A_{pp} & A_{pe} & B_p \\
A_{pe} & A_{ee} & B_e \\
B_p & B_e & D_\tau
\end{pmatrix},
\label{app:eq:M_components_matrix}
\end{equation}
where
\begin{equation}
\begin{aligned}
A_{pp}
&=
\frac{1}{\sigma_{l,p}^2}
+
\frac{K_e}{2}
+
2h_{\mathrm{ff}}^2\Lambda_p,
\\
A_{pe}
&=
\frac{K_e}{2}
+
h_{\mathrm{ff}}\Lambda_p,
\\
A_{ee}
&=
\frac{1}{\sigma_{l,e}^2}
+
\frac{K_e}{2}
+
\frac{\Lambda_p}{2},
\\
B_p
&=
-\frac{1}{\sigma_{l,p}^2}
+
2h_{\mathrm{ff}}g_{\mathrm{ff}}\Lambda_p,
\\
B_e
&=
g_{\mathrm{ff}}\Lambda_p,
\\
D_\tau
&=
\frac{1}{\sigma_{J,\tau}^2}
+
\frac{1}{\sigma_{l,p}^2}
+
2g_{\mathrm{ff}}^2\Lambda_p.
\end{aligned}
\label{app:eq:M_components}
\end{equation}
The negative term in $B_p$ follows from the cross term in $(\xi_p-\tau)^2$.

For completeness, the determinant entering the Gaussian integral can be
written explicitly as
\begin{equation}
\begin{aligned}
\det\mathbf M
&=
A_{pp}A_{ee}D_\tau
+
2A_{pe}B_pB_e
-
A_{pp}B_e^2
\\
&\quad
-
A_{ee}B_p^2
-
D_\tau A_{pe}^2.
\end{aligned}
\label{app:eq:detM_component_form}
\end{equation}
Although Eq.~\eqref{app:eq:detM_component_form} is already suitable for
direct numerical evaluation, its dependence on the Laplace variable $s$
can be made explicit.

We introduce the inverse squared longitudinal scales
\begin{equation}
\alpha_p
=
\frac{1}{\sigma_{l,p}^2},
\qquad
\alpha_e
=
\frac{1}{\sigma_{l,e}^2},
\qquad
\alpha_\tau
=
\frac{1}{\sigma_{J,\tau}^2}.
\label{app:eq:alpha_definitions}
\end{equation}
Using
\begin{equation}
K_e=s\kappa_e,
\qquad
\Lambda_p(s)
=
\frac{s\kappa_p}
{1+2s\kappa_p\sigma_{J,Z}^2},
\label{app:eq:Ke_Lambda_explicit}
\end{equation}
together with the identity
\begin{equation}
h_{\mathrm{ff}}+g_{\mathrm{ff}}=\frac{1}{2},
\label{app:eq:hg_identity}
\end{equation}
the determinant reduces to
\begin{equation}
\begin{aligned}
\det\mathbf M(s)
&=
D_0
+
s\kappa_e D_e
+
\Lambda_p(s)D_p
+
s\kappa_e\Lambda_p(s)D_{ep},
\end{aligned}
\label{app:eq:detM_explicit_s}
\end{equation}
where
\begin{equation}
D_0
=
\alpha_p\alpha_e\alpha_\tau,
\label{app:eq:D0_definition}
\end{equation}
\begin{equation}
D_e
=
\frac{1}{2}
\left(
\alpha_p\alpha_e
+
\alpha_p\alpha_\tau
+
\alpha_e\alpha_\tau
\right),
\label{app:eq:De_definition}
\end{equation}
\begin{equation}
D_p
=
\frac{1}{2}
\left(
\alpha_p\alpha_e
+
\alpha_p\alpha_\tau
+
4h_{\mathrm{ff}}^2
\alpha_e\alpha_\tau
\right),
\label{app:eq:Dp_definition}
\end{equation}
and
\begin{equation}
D_{ep}
=
g_{\mathrm{ff}}^2
\left(
\alpha_e+\alpha_\tau
\right).
\label{app:eq:Dep_definition}
\end{equation}
Thus, despite the three-dimensional Gaussian integration, the determinant
contains only terms proportional to
$1$, $s$, $\Lambda_p(s)$, and $s\Lambda_p(s)$.
Substituting the explicit expression for $\Lambda_p(s)$ gives
\begin{equation}
\begin{aligned}
\det\mathbf M(s)
&=
D_0
+
s\kappa_eD_e
\\
&\quad+
\frac{s\kappa_pD_p
+
s^2\kappa_e\kappa_pD_{ep}}
{1+2s\kappa_p\sigma_{J,Z}^2}.
\end{aligned}
\label{app:eq:detM_rational_s}
\end{equation}
Equation~\eqref{app:eq:detM_rational_s} displays the complete dependence
of $\det\mathbf M$ on the Laplace variable $s$. For positive rms widths and $s\geq0$, $\mathbf M(s)$ is positive definite, so the Gaussian integral is convergent.

The Gaussian integral over $\xi_p$, $\xi_e$, and $\tau$, including the normalization of $P_\tau$, is
\begin{equation}
\begin{aligned}
I_{\xi,\tau}
&=
\int_{-\infty}^{\infty}d\xi_p
\int_{-\infty}^{\infty}d\xi_e
\int_{-\infty}^{\infty}
\frac{d\tau}{\sqrt{2\pi}\sigma_{J,\tau}}
\exp\left[
-\frac{1}{2}
\mathbf q^{\mathsf T}\mathbf M\mathbf q
\right]
\\
&=
\frac{1}{\sqrt{2\pi}\sigma_{J,\tau}}
\frac{(2\pi)^{3/2}}
{\sqrt{\det\mathbf M}}
\\
&=
\frac{2\pi}
{\sqrt{\sigma_{J,\tau}^2\det\mathbf M}}.
\end{aligned}
\label{app:eq:gaussian_integral_matrix}
\end{equation}

Combining Eqs.~\eqref{app:eq:laplace_representation}, \eqref{app:eq:IZ_result}, and \eqref{app:eq:gaussian_integral_matrix}, the fully shot-averaged yield becomes
\begin{equation}
\begin{aligned}
\left\langle N_\gamma \right\rangle
&=
\frac{
\sigma_{\mathrm{eff}}N_eN_p
}
{
2\pi\sigma_{l,e}\sigma_{l,p}
}
\int_0^\infty ds\,
\\
&\quad\times
\frac{
\exp(-s\widetilde S_0)
}
{
\sqrt{1+2s\kappa_p\sigma_{J,Z}^2}
\sqrt{\sigma_{J,\tau}^2\det\mathbf M(s)}
}.
\end{aligned}
\label{app:eq:master_integral_matrix}
\end{equation}
This is the one-dimensional matrix form of the fully shot-averaged overlap integral.

\subsection{Explicit matrix-free form}

For direct numerical evaluation, it is convenient to combine the
focus-position and timing-jitter factors in the denominator of
Eq.~\eqref{app:eq:master_integral_matrix}. We define
\begin{equation}
\chi_Z
=
2\kappa_p\sigma_{J,Z}^2
\label{app:eq:chiZ_definition}
\end{equation}
and
\begin{equation}
\mathcal Q(s)
=
\sigma_{J,\tau}^2
\left(1+\chi_Zs\right)
\det\mathbf M(s).
\label{app:eq:Q_definition}
\end{equation}
Using Eq.~\eqref{app:eq:detM_rational_s}, this quantity becomes a
quadratic polynomial,
\begin{equation}
\mathcal Q(s)
=
q_0+q_1s+q_2s^2.
\label{app:eq:Q_polynomial}
\end{equation}
The coefficients are
\begin{equation}
q_0
=
\alpha_p\alpha_e,
\label{app:eq:q0_definition}
\end{equation}
\begin{equation}
\begin{aligned}
q_1
&=
\chi_Z\alpha_p\alpha_e
\\
&\quad+
\frac{\kappa_e}{2}
\left[
\sigma_{J,\tau}^2\alpha_p\alpha_e
+
\alpha_p+\alpha_e
\right]
\\
&\quad+
\frac{\kappa_p}{2}
\left[
\sigma_{J,\tau}^2\alpha_p\alpha_e
+
\alpha_p
+
4h_{\mathrm{ff}}^2\alpha_e
\right],
\end{aligned}
\label{app:eq:q1_definition}
\end{equation}
and
\begin{equation}
\begin{aligned}
q_2
&=
\frac{\chi_Z\kappa_e}{2}
\left[
\sigma_{J,\tau}^2\alpha_p\alpha_e
+
\alpha_p+\alpha_e
\right]
\\
&\quad+
\kappa_e\kappa_p g_{\mathrm{ff}}^2
\left(
1+\sigma_{J,\tau}^2\alpha_e
\right).
\end{aligned}
\label{app:eq:q2_definition}
\end{equation}
The polynomial representation remains finite in the limit
$\sigma_{J,\tau}\rightarrow0$, even though
$\alpha_\tau=\sigma_{J,\tau}^{-2}$ diverges separately.

Substitution into Eq.~\eqref{app:eq:master_integral_matrix} gives the
explicit matrix-free master integral
\begin{equation}
\boxed{
\begin{aligned}
\left\langle N_\gamma \right\rangle
&=
\frac{
\sigma_{\mathrm{eff}}N_eN_p
}
{
2\pi\sigma_{l,e}\sigma_{l,p}
}
\int_0^\infty ds\,
\frac{
\exp(-s\widetilde S_0)
}
{
\sqrt{q_0+q_1s+q_2s^2}
}.
\end{aligned}
}
\label{app:eq:master_integral_polynomial}
\end{equation}
All dependence on the Laplace variable $s$ is therefore contained in
the exponential $\exp(-s\widetilde S_0)$ and in the quadratic polynomial
$\mathcal Q(s)=q_0+q_1s+q_2s^2$. Equation~\eqref{app:eq:master_integral_polynomial}
is the form used for the numerical parameter scans.

\subsection{Equivalent Schur-complement form}

For numerical implementation it is sometimes convenient to perform the timing-jitter integral first and reduce the remaining integral over $\xi_p$ and $\xi_e$ to a two-dimensional Gaussian. Starting from Eq.~\eqref{app:eq:M_components_matrix}, the integration over $\tau$ gives the Schur-complement coefficients
\begin{equation}
\begin{aligned}
\overline A_{pp}
&=
A_{pp}
-
\frac{B_p^2}{D_\tau},
\\
\overline A_{pe}
&=
A_{pe}
-
\frac{B_pB_e}{D_\tau},
\\
\overline A_{ee}
&=
A_{ee}
-
\frac{B_e^2}{D_\tau}.
\end{aligned}
\label{app:eq:Schur_coefficients}
\end{equation}
The timing integral contributes the factor
\begin{equation}
\frac{1}{\sqrt{\sigma_{J,\tau}^2D_\tau}},
\label{app:eq:timing_normalization}
\end{equation}
and the remaining two-dimensional Gaussian integral gives
\begin{equation}
\int d\xi_p\,d\xi_e\,
\exp\left[
-\frac{1}{2}
\left(
\overline A_{pp}\xi_p^2
+
2\overline A_{pe}\xi_p\xi_e
+
\overline A_{ee}\xi_e^2
\right)
\right]
=
\frac{2\pi}
{
\sqrt{
\overline A_{pp}\overline A_{ee}
-
\overline A_{pe}^2
}
}.
\label{app:eq:two_dimensional_gaussian}
\end{equation}
Thus Eq.~\eqref{app:eq:master_integral_matrix} is equivalently
\begin{equation}
\begin{aligned}
\left\langle N_\gamma \right\rangle
&=
\frac{
\sigma_{\mathrm{eff}}N_eN_p
}
{
2\pi\sigma_{l,e}\sigma_{l,p}
}
\int_0^\infty ds\,
\\
&\quad\times
\frac{
\exp(-s\widetilde S_0)
}
{
\sqrt{1+2K_p\sigma_{J,Z}^2}
\sqrt{\sigma_{J,\tau}^2D_\tau}
\sqrt{
\overline A_{pp}\overline A_{ee}
-
\overline A_{pe}^2
}
}.
\end{aligned}
\label{app:eq:master_integral_schur}
\end{equation}
The equivalence follows from
\begin{equation}
\det\mathbf M
=
D_\tau
\left(
\overline A_{pp}\overline A_{ee}
-
\overline A_{pe}^2
\right).
\label{app:eq:det_schur_identity}
\end{equation}

\subsection{Useful limiting cases}

First, consider the limit of zero laser arrival-time jitter,
\begin{equation}
\sigma_{J,t}\rightarrow0,
\qquad
\sigma_{J,\tau}=c\sigma_{J,t}\rightarrow0.
\label{app:eq:no_timing_limit}
\end{equation}
In this limit
\begin{equation}
D_\tau\rightarrow\infty,
\qquad
\sigma_{J,\tau}^2D_\tau\rightarrow1,
\qquad
\overline A_{ij}\rightarrow A_{ij}.
\label{app:eq:no_timing_coefficients}
\end{equation}
Therefore Eq.~\eqref{app:eq:master_integral_schur} reduces to
\begin{equation}
\begin{aligned}
\left\langle N_\gamma \right\rangle_{\sigma_{J,t}=0}
&=
\frac{
\sigma_{\mathrm{eff}}N_eN_p
}
{
2\pi\sigma_{l,e}\sigma_{l,p}
}
\int_0^\infty ds\,
\\
&\quad\times
\frac{
\exp(-s\widetilde S_0)
}
{
\sqrt{1+2K_p\sigma_{J,Z}^2}
\sqrt{
A_{pp}A_{ee}
-
A_{pe}^2
}
}.
\end{aligned}
\label{app:eq:master_integral_no_timing}
\end{equation}

Second, if the longitudinal focus-position jitter is zero,
\begin{equation}
\sigma_{J,Z}=0,
\label{app:eq:no_Z_jitter_condition}
\end{equation}
then
\begin{equation}
\Lambda_p=K_p,
\qquad
\sqrt{1+2K_p\sigma_{J,Z}^2}=1.
\label{app:eq:no_Z_jitter_limit}
\end{equation}

Third, for a stationary Gaussian waist,
\begin{equation}
\beta_{\mathrm{ff}}=0,
\qquad
h_{\mathrm{ff}}=\frac{1}{2},
\qquad
g_{\mathrm{ff}}=0.
\label{app:eq:stationary_waist_limit}
\end{equation}
The timing error then shifts only the longitudinal laser envelope through $\xi_p-\tau$ and does not enter the waist coordinate $\zeta=z-\Delta Z$.

Finally, if both diffraction and electron beta-function evolution are neglected,
\begin{equation}
\kappa_e=0,
\qquad
\kappa_p=0,
\label{app:eq:no_diffraction_no_divergence}
\end{equation}
then $\widetilde S=\widetilde S_0$ is constant. The longitudinal integrations are normalized and independent of the jitter variables, giving
\begin{equation}
\left\langle N_\gamma \right\rangle
=
\frac{
\sigma_{\mathrm{eff}}N_eN_p
}
{
2\pi
\left(
\sigma_{e,0}^2+
\sigma_{p,0}^2+
\sigma_{J,\perp}^2
\right)
}.
\label{app:eq:constant_width_limit}
\end{equation}
This limit provides a useful check: when the transverse widths do not vary along the collision, longitudinal focus-position jitter and arrival-time jitter do not change the total four-dimensional overlap.

%
\end{document}